\documentclass[preprints,review,accept,moreauthors,pdftex,10pt,a4paper]{mdpi}

\def\mnras{{\it Mon.~Not. R. Astron.~Soc.}}

\def\aap{{\it Astron. Astrophys.}}

\def\nat{{\it Nature}}

\def\apss{{\it Astrophys. Space Sci.}}

\newcommand\simlt{\lower.5ex\hbox{$\; \buildrel < \over \sim \;$}}
\newcommand\simgt{\lower.5ex\hbox{$\; \buildrel > \over \sim \;$}}

\firstpage{1}
\pubvolume{xx}
\issuenum{1}
\articlenumber{1}
\pubyear{2019}
\copyrightyear{2019}
\history{Received: 17 December 2018; Accepted: 24 January 2019; Published: date}

\pdfoutput=1

\Title{Gamma-Ray Astrophysics in the Time Domain}

\Author{Frank M. Rieger \orcidA{}}

\AuthorNames{Frank M. Rieger}

\address[1]{%
ZAH, Institut f\"ur Theoretische Astrophysik, Heidelberg University, Philosophenweg 12,
69120 Heidelberg, Germany; f.rieger@uni-heidelberg.de}

\abstract{The last few years have seen gamma-ray astronomy maturing and advancing in the field of
time-domain astronomy, utilizing source variability on timescales over many orders of magnitudes, from a
decade down to a few minutes and shorter, depending on the source. This review focuses on some of the
key science issues and conceptual developments concerning the timing characteristics of active galactic
nuclei (AGN) at gamma-ray energies. It highlights the relevance of adequate statistical tools and illustrates
that the developments in the gamma-ray domain bear the potential to fundamentally deepen our
understanding of the nature of the emitting source and the link between accretion dynamics, black hole
physics, and jet ejection.}
\keyword{gamma-rays; emission: non-thermal; variability; origin: jet; origin: black hole}

\begin{document}
\section{Introduction}
The last decade has seen tremendous experimental progress in gamma-ray astronomy much beyond simple source
detection (e.g., see \cite{Rieger2013,Funk2015,Madejski2016} for a review). In many cases, detailed spectral and timing
characterization have become possible allowing one to probe deeply into the nature and physics of these sources. In
particular, gamma-ray astronomy has by now matured and progressed further in the field of time-domain astronomy,
utilizing source variability on timescales over many orders of magnitudes, from a decade down to a few minutes
and shorter. Instruments such as the Fermi Large Area Telescope ({Fermi}-LAT), the High-Altitude Water
Cherenkov Gamma-Ray Observatory (HAWC) or the First G-APD Cherenkov Telescope (FACT), for example, have
opened up the possibility for unbiased long-term (timescales up to several years) studies of bright astrophysical
objects in the high energy (HE; $>$100 MeV) and the very high energy (VHE; $>$100 GeV) domain, respectively
(e.g., \cite{Thompson2018,HAWC2017,FACT2017}), while modern Imaging Atmospheric Cherenkov Telescopes
(IACTs) have demonstrated their excellent capabilities to characterize VHE flaring states down to below sub-hour
timescales.

This paper focuses on some of the key issues and conceptual developments concerning the timing characteristics of
AGN (including rapid variability, log-normal flux distributions, power-law noise, and quasi-periodic oscillations) at
gamma-ray energies.

\section{Timing Characteristics}
\subsection{Rapid Variability at VHE Energies}

\begin{figure}[H]
 \centering
 \includegraphics[width=270pt]{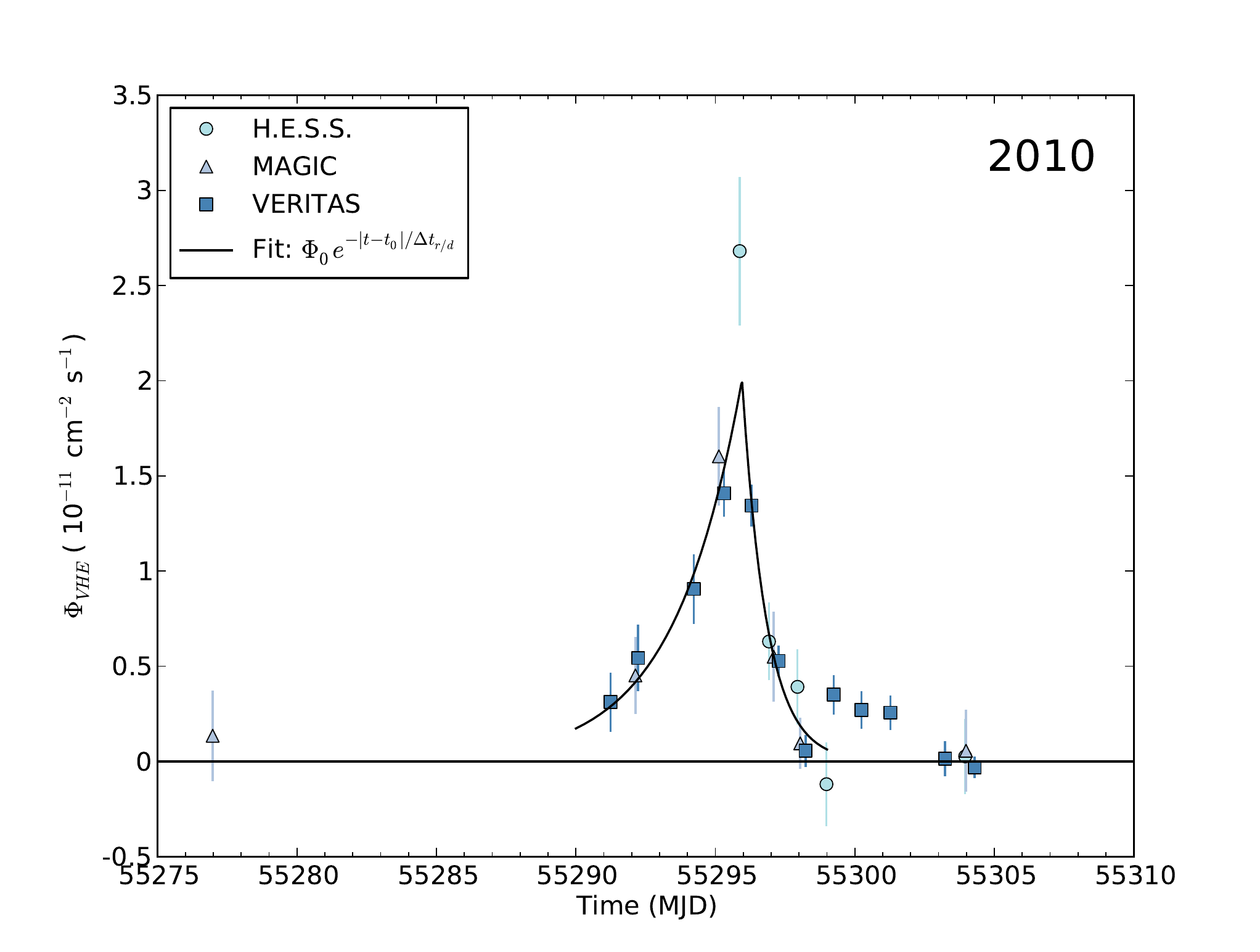}
 \caption{The very high energy (VHE) flare of the radio galaxy M87 in April 2010 as observed by H.E.S.S., MAGIC and
 VERITAS.
 Significant day-scale activity is evident. Since the jet in M87 is believed to be misaligned (suggestive of modest
 Doppler beaming only) and its black hole light-crossing scale large ($r_s/c\simgt 0.5$ days), this VHE activity is
 extremely fast. The curve represents a fit by an exponential function, indicating doubling times of $\sim 1.7$ days
 and $\sim 0.6$ days during rise and decay, respectively. From~\cite{Abramowski2012}.}\label{M87_VHE}
\end{figure}

\begin{figure}[H]
 \centering
 \includegraphics[width=350pt]{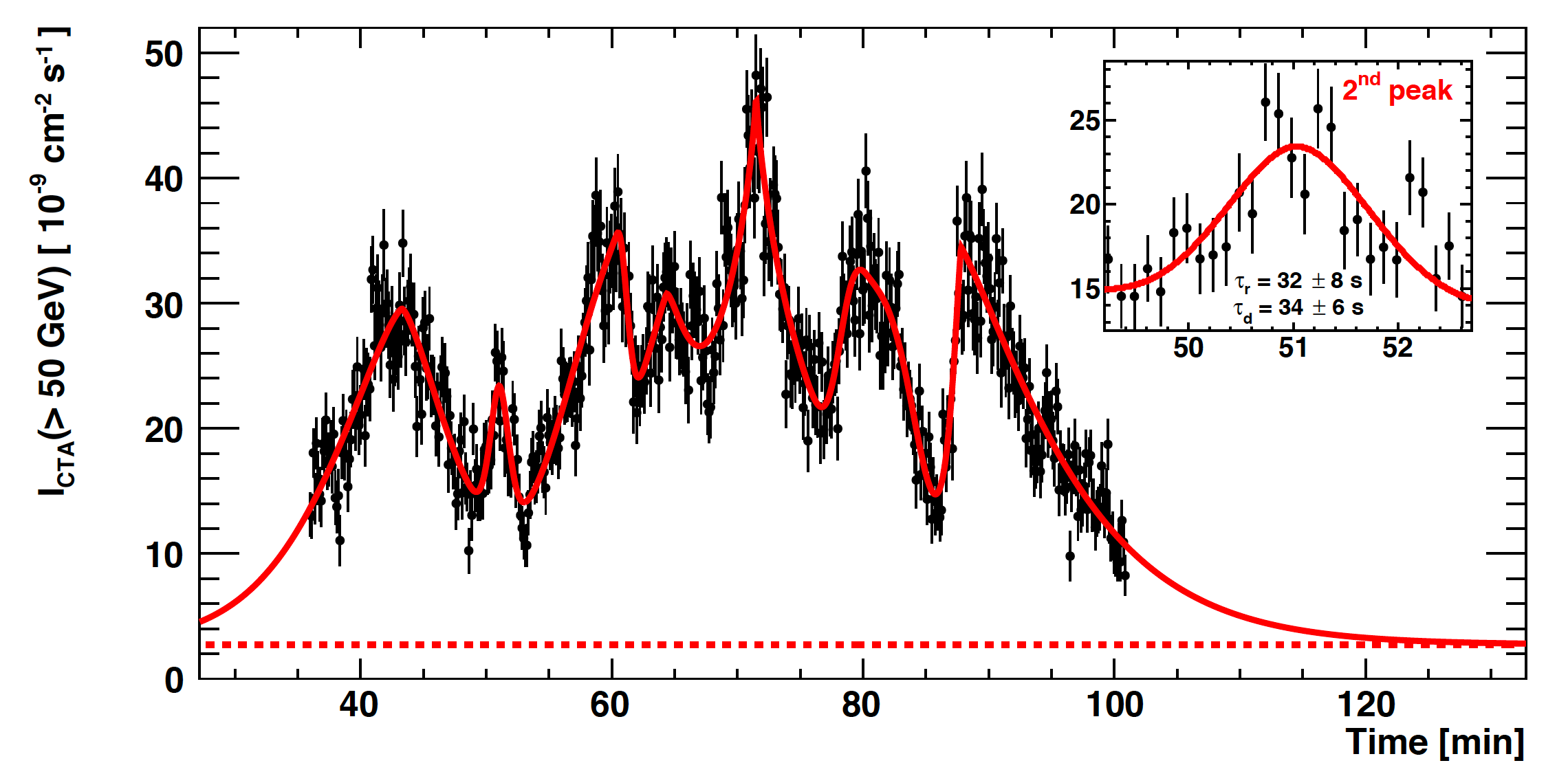}
 \caption{Simulated CTA light curve based on an extrapolation of the power spectrum for the strong 2006 VHE flare
 of the blazar PKS 2155-304. CTA will allow probing sub-minute timescales (see inlay) and thereby help to discriminate
 between different physical models. From~\cite{CTA2017}.}\label{PKS2155_CTA}
\end{figure}

Radio-loud AGN can be highly variable gamma-ray emitters. This not only applies to the strongly-Doppler-boosted blazar
sources, but also to misaligned jet sources such as radio galaxies. VHE doubling timescales down to a few minutes ($\Delta
t_{\rm obs} \sim$ 2--3 min) are known for the VHE blazars Mkn 501 \cite{Albert2007} and PKS 2155-304 \cite{Aharonian2007}
and at HE for the flat spectrum radio quasar (FSRQ) blazar 3C279 \cite{Ackermann2016}. Interestingly, rapid VHE activity has 
been seen in radio galaxies as well, i.e., from day-scale for M87 \cite{Aharonian2006,Abramowski2012}, see Figure~\ref{M87_VHE}, 
to~intra-day ($\Delta t_{\rm obs}\sim 10$ h) for NGC 1275 \cite{Ansoldi2018}, down to minute-scale for IC~310 \cite{Aleksic2014a}; 
see~\cite{Rieger2018}. Rapid variability is usually taken to indicate extreme jet conditions. Causality arguments, for example,
imply that the emission would need to arise in a very compact region of comoving size $\Delta r'
\simlt \delta c \Delta t_{\rm obs} = 10^{15} (\delta/10) (\Delta t_{\rm obs}/1\,\mathrm{h})$ cm (where $\delta$ is the Doppler
factor), possibly close to the black hole at distances $d \simlt 2 c\gamma_b^2 \Delta t_{\rm obs} \simeq 2\times 10^{16}
(\gamma_b/10)^2 (\Delta t _{\rm obs}/1\,\mathrm{h})$ cm (where $\gamma_b\geq 1$ is the jet bulk Lorentz factor) only, i.e.,
typically less than a few hundred Schwarzschild radii away. Despite its compactness, the emitting region often needs to
be highly luminous to account for the observed gamma-ray output, and this can impose a severe constraint on possible
models (see below). In some cases such as for PKS 2155-304, the VHE light curve during the high state is sufficiently resolved
to suggest the presence of multiple (possibly interacting) emitting zones. The discriminating potential will strongly increase
with the upcoming Cherenkov Telescope Array (CTA). Figure~\ref{PKS2155_CTA} shows a simulation based on the strong 
VHE flare in 2006, demonstrating the capability to probe sub-minute timescales.
\unskip
%

In the extragalactic jetted-source context, rapid variability comes along with a general challenge. In~the conventional picture
of an emission region embedded in a uni-directional jet flow, the observed variability imposes a constraint on the size of the
central engine that is independent of Doppler boosting (e.g., \cite{Levinson2008,Dermer2008}). To see this, note that the
observed timescale $\Delta t_{\rm obs}$ is related to the comoving (flow) timescale $\Delta t'$ by $\Delta t_{\rm obs} =(1+z)
\Delta t'/\delta$, where $\delta=1/(\gamma_b [1-\beta\cos i])$ is the Doppler factor, and that by causality $\Delta t' \simgt
\Delta r'/c$. Length scales in the rest frame of the central engine and the comoving (flow) frame, on the other hand, are related
by $\Delta r = \Delta r'/\gamma_b$ (length contraction). Noting that the characteristic length scale in the rest frame of the
central engine driving the jet is $\Delta r \sim r_s \equiv 2 GM_{BH}/c^2$, one finds for the generic variability timescale:
\begin{equation}\label{BH_size}
\Delta t_{\rm obs} \simgt \frac{(1+z)}{c}\, \frac{\gamma_b}{\delta}\, r_s \simgt \frac{(1+z)}{2} \,\frac{r_s}{c}\,,
\end{equation} using that $\delta=2\gamma_b$ in the head-on ($i=0$) approximation. The $\sim3$-min VHE variability
observed in PKS 2155-304 ($z=0.116$) would then imply a black hole mass of $M_{BH} \simlt 4 \times 10^7 M_{\odot}$ only,
much smaller than what is inferred from the host-galaxy luminosity relation (e.g., \cite{Rieger2010}). Faster variability could
only be achieved if the jet collided with a small obstacle (size $r_c \ll r_s$). Yet, in this case, the jet power that could be
tapped for producing radiation is much smaller, roughly by a factor $(r_c/r_s)^2$, so that typically, very high jet powers would
be needed to account for the observed VHE emission.

\begin{figure}[H]
\centering
\includegraphics[angle=0, width=5.6cm]{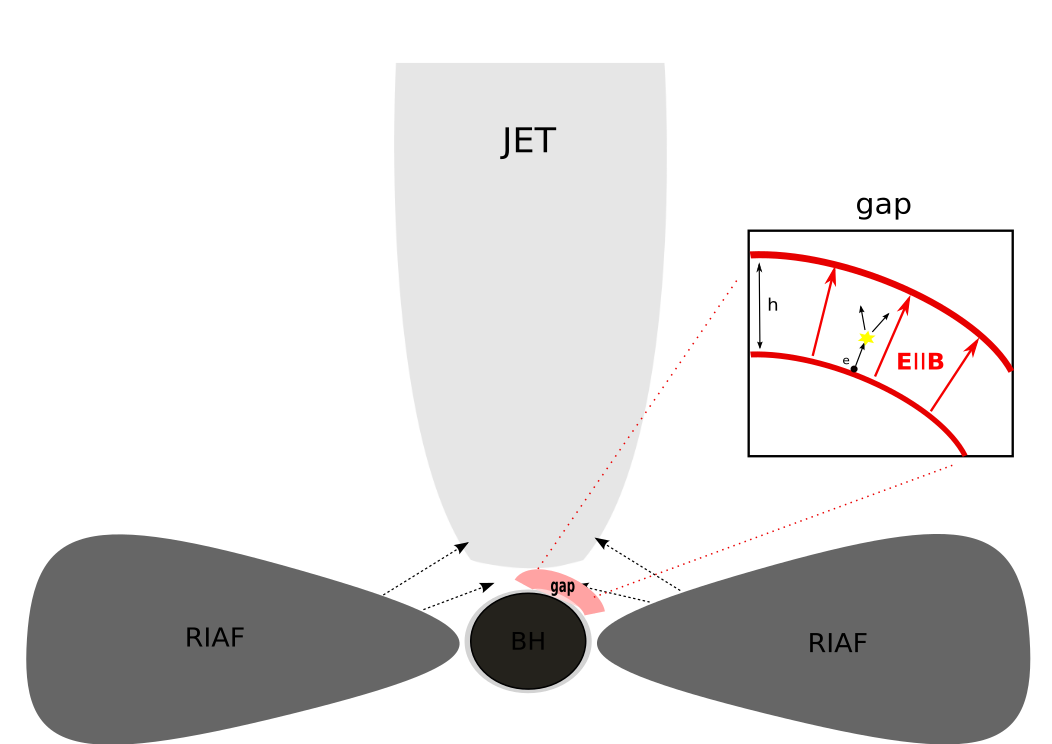}
\hspace{0.1cm}
\includegraphics[angle=0, width=4.8cm]{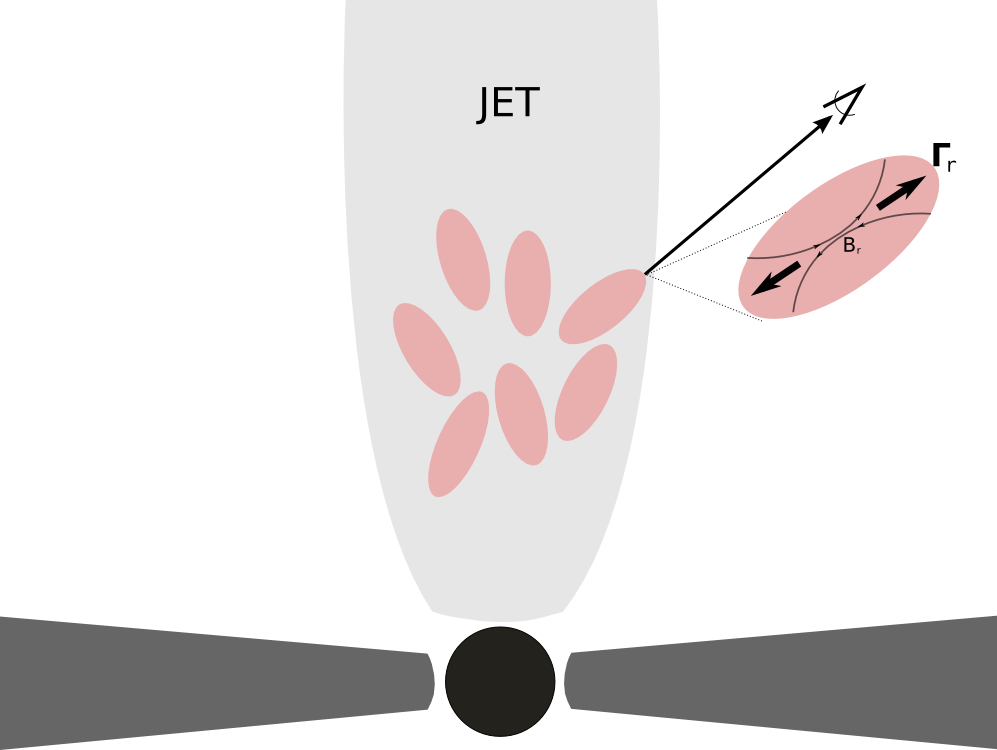}
\hspace{0.1cm}
\includegraphics[angle=0, width=4.5cm]{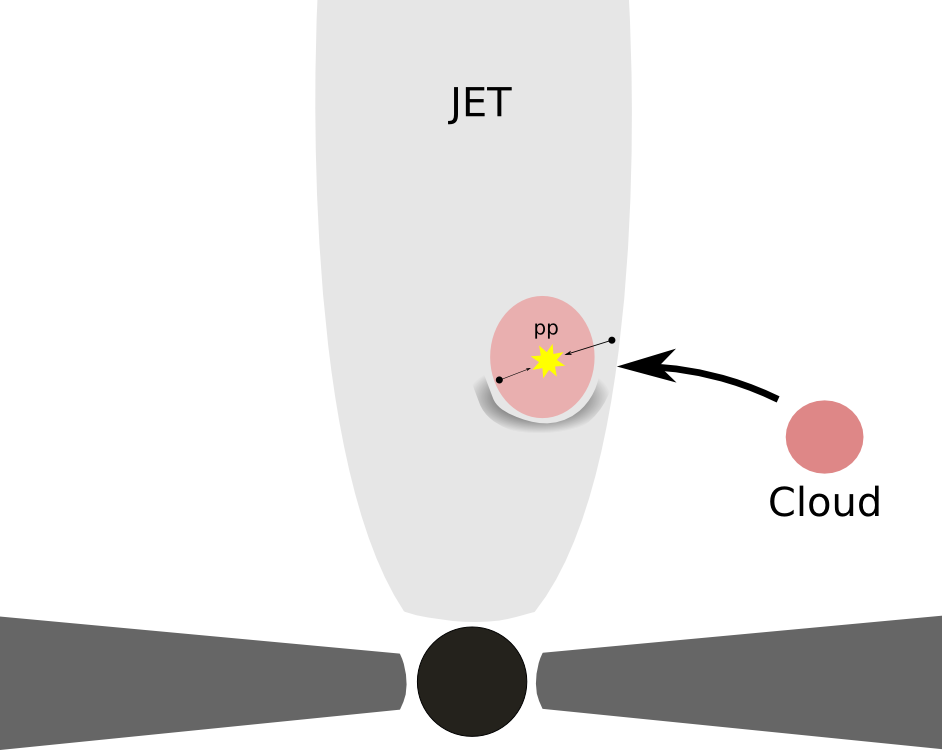}
\caption{Possible scenarios for the origin of rapidly-variable VHE emission in AGN. Left: Particle acceleration in unscreened
(``gap'') electric fields close to the black hole horizon triggers an electron-positron pair cascade, which is accompanied by inverse
Compton up-scattering of the ambient disk photon to the VHE regime. VHE variability then becomes a possible signature of jet
formation~\citep{Levinson2011}. Middle: Jets-in-jet model \cite{Giannios2009}, where a variety of ``mini-jet'' features (``plasmoids'')
is produced by magnetic reconnection events within the main jet flow. This could lead to an additional velocity component
relative to the main flow ($\Gamma_r$) and allow for a favorable orientation (i.e., increased Doppler boosting and time scale
reduction) with respect to the observer. Right: Illustration of a hadronic model, where interactions of the jet with a small, massive
obstacle (star or cloud) facilitate shock-acceleration and introduce a sufficient target density to allow for efficient $pp$-collisions
\cite{Barkov2012}.}
\label{models}
\end{figure}

In principle, however, the dynamics and the geometry of the emission region(s) could be much more complex. Salvati et
al.~\cite{Spada1998}, for example, have argued that if the emission is due to a series of conical shocks, fast blazar variability
does not necessarily require extreme jet parameters. In addition the observational findings have triggered new conceptual
developments in which rapid variability is related to, e.g., moving subregions or turbulence within the jet, black hole
magnetospheric processes, or the propagation of non-linear electromagnetic waves (e.g., \cite{Begelman2008,Ghisellini2009,
Kirk2011,Narayan2012,Hirotani2016,Aharonian2017}). Major reference scenarios currently include black hole horizon gaps,
jets-in-jet, or jet interaction models (e.g., \cite{Levinson2011,Giannios2009,Barkov2012}); see Figure~\ref{models} for an illustration.
Each of these has its own challenge from, e.g., limited power extraction in the case of magnetospheric gaps and unusual
magnetizations in the case of reconnection-induced mini-jets to the requirement of maximal jet powers in the case of jet
interaction models (see~\cite{Rieger2018} for a recent discussion). At the current stage, the most promising way to gain
further insights into the physical origin of variability seems to be to go beyond classical minimum variability considerations and
homogeneous one-zone approaches and to take into account the full timing characteristics of the emission, as outlined below.
Methodologically, this draws on the general
concept that the variability in AGN can be fairly well understood as a stochastic process. An observed time series (i.e., a
light curve) is then seen as a realization of the underlying stochastic process that is sampled by the observation.\footnote{There
are two common approaches to deal with irregular sampling in stochastic light curves, i.e., by means of extensive Monte Carlo
simulations of artificial light curves \cite{Timmer1995,Emma2013} or by means of likelihood-based approaches using special
parameterized stochastic models in the time domain such as, e.g., the first-order continuous-time autoregressive (Ornstein--Uhlenbeck)
process (e.g., \cite{Sobolewska2014,Kelly2014}). The latter offer an efficient means to extract information from large time-domain
datasets, in particular the power spectral density (PSD). While autoregressive models are becoming increasingly popular (cf.~\cite
{Feigelson2018,Moreno2018}), general caveats concern the limitations of linear (e.g., ARMA, CARMA) models to extract physical
information from (very probably) non-linear systems (e.g., \cite{Vio2005}); see also Section~\ref{lognormal}.}

\subsection{PDF Shape and Log-Normality}\label{lognormal}

\begin{figure}[H]
\centering
\includegraphics[width=350pt]{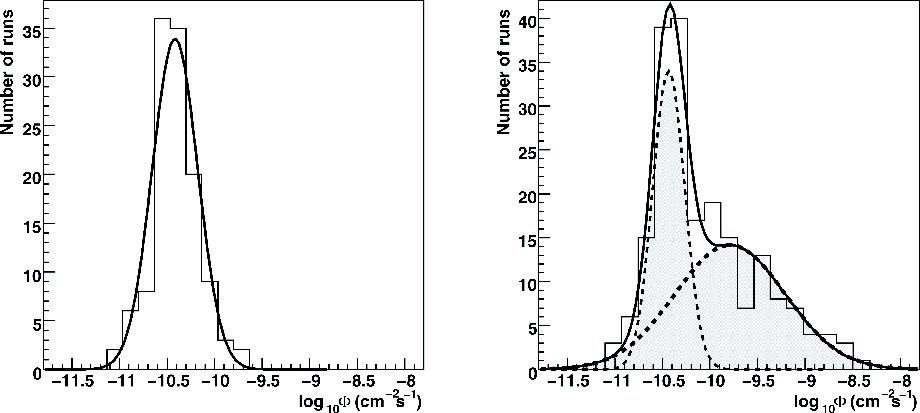}
\caption{Distributions of the logarithms of integral VHE fluxes above $200~{\rm GeV}$ for PKS 2155-304.
Left: From 2005--2007, without the strong flare period in July 2006. Right: All data, from 2005--2007; the solid
line represents the fit by the sum of two Gaussians. The findings provide evidence that the source switches from
a quiescent VHE state with rather minimal activity to a flaring state, with the flux distribution in each state
following a lognormal distribution (see also \cite{Abdalla2017}). From~\cite{pks2155_2010}.}
\label{pks2155log}
\end{figure}

\begin{figure}[H]
\centering
\includegraphics[width=190pt]{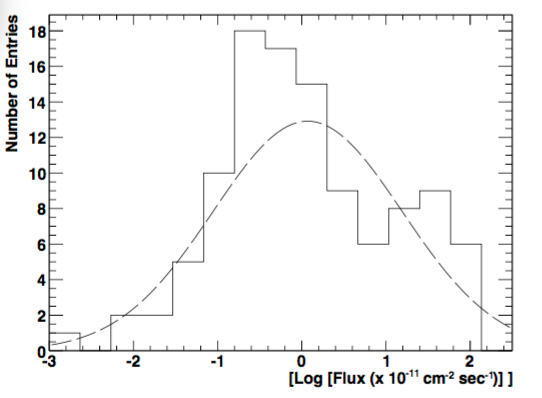}
\caption{Distributions of VHE fluxes ($\geq$2 TeV) for the flare of Mkn 501 during 19--24 June 2014 based
on a four-minute time binning. Furthermore, in this case, a log-normal distribution (dashed curve) is preferred over a
Gaussian. The difference is not as significant, though (see also~\cite{Romoli2018}), and one could also
speculate whether the sum of two Gaussians might provide a better characterization of the log(flux)-distribution.
From~\cite{mkn501_lognormal}.}
\label{mkn501log}
\end{figure}

Important insights can be obtained by characterizing the underlying process driving the variability by its observed probability
density function (PDF). The PDF can in principle be estimated by fitting a model function to the histogrammed data. As
a simple comparison, one can study the distribution of VHE fluxes (e.g., the number of runs as a function of the measured
photon flux or the logarithm thereof) and evaluate the appropriateness of a fit by a Gaussian distribution as representative
of a normal random process. Figures~\ref{pks2155log} and \ref{mkn501log} provide examples in the case of PKS 21455-304
and Mkn 501, where a clear preference for the logarithm of the flux to be Gaussian (normal) distributed, i.e., for log-normality,
has been found \cite{pks2155_2010,mkn501_lognormal}.\footnote{A random variable $X$ is log-normally distributed if $\log(x)$
obeys a normal (Gaussian) distribution. The PDF of such a random variable is of the form $f(x) = \frac{1}{x\,\sigma \sqrt{2\pi}}
\exp\left(-\frac{1}{2\sigma^2}(\log[x]-\mu)^2\right)$, where $\mu$ is the mean and $\sigma$ the standard deviation.}
Similar results are obtained by studying the relationship between the mean flux and the absolute $rms$ (root mean square)
variability (e.g., \cite{Chevalier2015,Abdalla2017}),\footnote{Defined as the square-root of the light curve variance and related
to the square-root of the integral of the PSD, see below, over the observable frequency range, i.e., $\left(\int PSD(\nu) \cdot
d\nu\right)^{1/2}$.} where a linear relationship is known to provide evidence for a log-normal distribution of the
fluxes \cite{Uttley2001,Uttley2005}.

\unskip

Where statistics permits, current evidence suggests that the gamma-ray fluxes in blazars are preferentially log-normally
distributed (cf. also~\cite{Sinha2016} in the case of Mkn 421 and~\cite{Shah2018} for the case of bright Fermi
blazars). Lognormal flux variability is not unusual, but in fact is well-known for accreting galactic sources such as X-ray
binaries \cite{Uttley2001}, where it has been linked to the underlying accretion process. The current findings are of interest
since in a lognormal process, the fluctuations of the flux are on average proportional, or at least correlated, to the flux itself.
This rules out additive processes for the origin of the observed variability and instead favors multiplicative ones.\footnote{For 
a stationary stochastic process $X$ that results from a multiplication of $N$ random subprocesses $x_i$, $X =
\prod x_i$, 
the logarithm of $X$ is equivalent to the sum of the logarithm of the individual $x_i$, i.e., $\log X =
 \log x_1 + \log x_2 +...+\log x_N$. By~the central limit theorem, this sum must approach a normal (Gaussian) distribution for
 $N \rightarrow \infty$. Astrophysically, $N$ does not have to be large to achieve a good log-normal distribution \cite{Ioka2002}.}
If this is true, then additive models (e.g., shot-noise or a simple superposition of many ``mini-jets'') are no longer adequate
to describe the observed variability behavior, and multiplicative, cascade-like scenarios need to be invoked.

\begin{figure}[H]
 \centering
 \includegraphics[width=300pt]{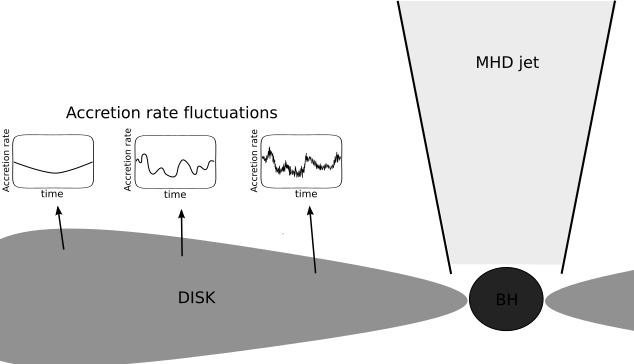}
 \caption{A possible accretion-disk origin for log-normality. Density fluctuations in the disk, arising on local viscous
 timescales, can propagate inwards and couple together so as to produce a multiplicative behavior in the accretion
 rate. If this is transmitted to the jet and picked up by the acceleration mechanism, the resultant VHE emission might
 share in the underlying log-normal characteristics.}\label{lognormal_model}
\end{figure}

In principle, several scenarios for the origin of log-normal flux distributions could be envisaged:

(i) Similar as for X-binaries, the inferred log-normality at gamma-ray energies in blazars could mark the influence of the
accretion disk on the jet (e.g., \cite{Giebels2009,McHardy2010}). Independent density fluctuations in the accretion disk
on local viscous timescale, i.e., $t_v(r) \sim (1/\alpha)\,(r/h)^2 (r/r_g)^{3/2} r_g/c$ in the case of a standard disk (where $r_g:=
GM_{BH}/c^2$, $h$ is the disk height, and $\alpha$ the viscosity parameter), provide one possible realization. Provided
damping is negligible, these fluctuations can propagate inward and couple together to produce a multiplicative (power-law
noise) behavior in the innermost disk part~\cite{Lyubarski1997,King2004, Arevalo2006}. If~this behavior is efficiently
transmitted to the jet (particle injection rate), the gamma-ray emission could be modulated accordingly \cite{Rieger2010};
see Figure~\ref{lognormal_model}. This requires, amongst others, that the timescale for particle acceleration and radiative
losses within the jet be correspondingly small~\cite{Rieger2010}.
Log-normality at different wavelengths and over long timescales might then well be feasible (i.e., up to timescales
$t_v(r_d)$ where $r_d$ is the characteristic disk radius). Short timescales (on horizon light-crossing times), however,
could only be achieved in the case of a thick (inner) disk (where $h\sim r$); see also below.\\

(ii) Cascade-related emission processes (e.g., proton-induced synchrotron cascades, cf. \cite{Mannheim1993}, or
magnetospheric inverse-Compton pair production cascades, cf. \cite{Levinson2011}) might possibly lead to some
log-normal flux distributions. Relevant constraints in these cases arise, however, from the energy bands in which
log-normality has been seen (e.g., in the optical, X-ray, and gamma-ray range for PKS 2155-304 \cite{Chevalier2015})
and the timescales over which log-normality has been found (i.e., from sub-hour to yearly timescales at gamma-ray
energies). The latter are expected to be limited by the gap travel time for magnetospheric processes and the dynamical
or escape time for hadronic cascades.

\begin{figure}[H]
 \centering
 \includegraphics[width=220pt]{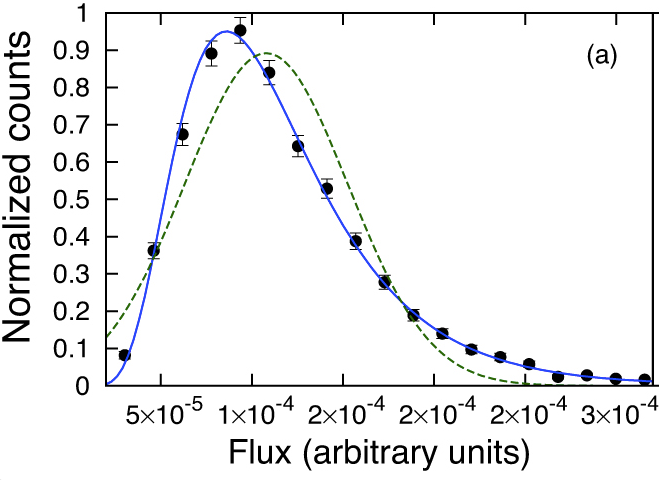}
 \caption{Log-normality in the accelerated particle number density arising from random fluctuations in the particle acceleration
 rate at a shock front. The figure shows the histogram of the particle number density at $\gamma=10^3$ based on simulations
 in which Gaussian perturbations are introduced into $t_{\rm acc}$. The~dashed green line represents the best-fitting Gaussian
 and the solid blue line the best-fitting log-normal PDF. Particle injection with $\gamma_0=10$ has been assumed. 
 From~\cite{Sinha2018}.}\label{log_acc} \end{figure}

(iii) Alternatively, log-normality could be associated with the acceleration process itself, e.g., with random fluctuations
in the particle acceleration rate \cite{Sinha2018}. In the case of diffusive shock acceleration, for example, the accelerated
particle distribution contains powers in $t_{\rm acc}$, i.e., $n(\gamma) \propto t_{\rm acc} \gamma^{-1-t_{\rm acc}/
t_{\rm esc}} (1-\gamma/\gamma_{\rm max})^{t_{\rm acc}/t_{\rm esc}-1}\,(1/\gamma_0-1/\gamma_{\rm max})^{-t_{\rm acc}/
t_{\rm esc}}$, where $t_{\rm acc} \sim \kappa/u_s^2$ denotes the acceleration and $t_{\rm esc}$ the escape timescale,
respectively \cite{Kirk1998}, $\gamma_0$ is the injection Lorentz factor, $u_s$ is the shock speed, and $\gamma_{\rm max}$
is the maximum Lorentz factor determined by radiative losses.
If diffusion $\kappa$ is characterized by Gaussian perturbations such that $t_{\rm acc} = t_{\rm acc,0}+\Delta t_{\rm acc}$,
then the fractional variability $\Delta n(\gamma)/n(\gamma)$ becomes a linear combination of Gaussian and
$\log(\gamma)$-terms \cite{Sinha2018}. Depending on the energy scale, the accelerated particle number density could
thus resemble a lognormal distribution. Figure~\ref{log_acc} provides an illustration of this effect. Log-normality in the
radiating particle distribution could then well result in a log-normal flux distribution in the case of synchrotron or external
inverse Compton. Log-normality in this case, however, would be energy-dependent, with its significance becoming weaker
towards lower energies and disappearing for energies close to a threshold ($\gamma \rightarrow \gamma_0$).

PDF studies at different frequencies (cf. \cite{Chevalier2019} for PKS 2155-304) could be of help to distinguish between the
noted scenarios. In principle, the above does not preclude the possibility that log-normality in different states (long term, low
level, and short term, flare) arises due to different processes.

\subsection{PSD and Power-Law Noise}
In time series analysis, the power spectral density (PSD) represents an important tool for characterizing variability. The PSD
provides a measure for the contribution of different timescales to the variability power, i.e., it quantifies the amount of variability
power as a function of (temporal) frequency ($\nu\sim 1/t$). Methodologically, this implies a move from a signal description in
the time domain to one in the frequency domain by means of Fourier transformation. In practice, we can only get an estimate
of the PSD, and the simplest method is to estimate the PSD by the periodogram (i.e., the squared modulus of the discrete
Fourier Transform, properly normalized). Accordingly, a periodicity at a given timescale would lead to a peak in the PSD at
the corresponding frequency, while a continuum power spectrum would correspond to non-periodic signals. The latter are
typically parameterized as power-law noise with a PSD of the form:
\begin{equation}
PSD(\nu) \propto \nu^{-\beta}\,,
\end{equation} where $\beta\geq0$ is
the power index (e.g., $\beta=0$ for white noise, $\beta=1$ for pink or flicker noise, and $\beta=2$ for red or Brownian noise).
A broken power-law PSD would then be indicative of a characteristic timescale ($t_b$) at the corresponding break frequency
($\nu_b\sim 1/t_b$).

\begin{figure}[H]
 \centering
 \includegraphics[width=450pt]{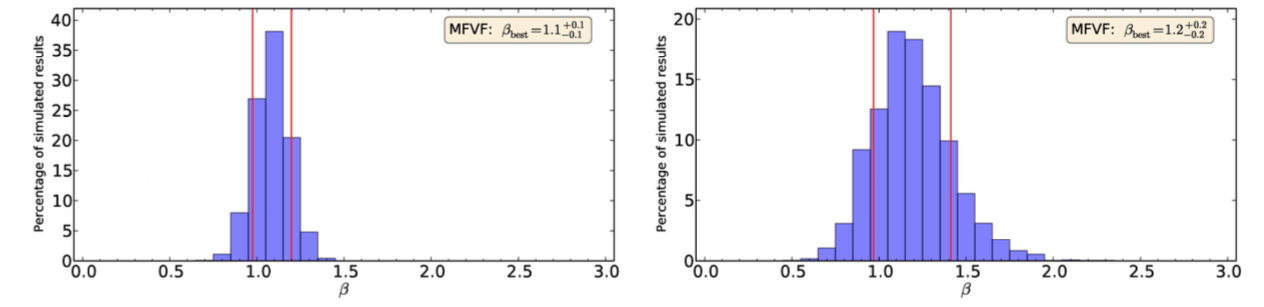}
 \caption{Best-fit parameters and uncertainties for the power-law noise index $\beta$ for the long-term gamma-ray
 activity in PKS 2155-304 estimated from simulated light curves. Left: For VHE (H.E.S.S.) data. Right: For HE
 ({Fermi}-LAT) data. The vertical red bars indicate the 1$\sigma$ uncertainties on the best fit. Both the VHE and
 the HE variability are compatible with flicker noise $\beta =1$. From~\cite{Abdalla2017}.}\label{PKS2155_PSD}
\end{figure}

\begin{figure}[H]
 \centering
 \includegraphics[width=220pt]{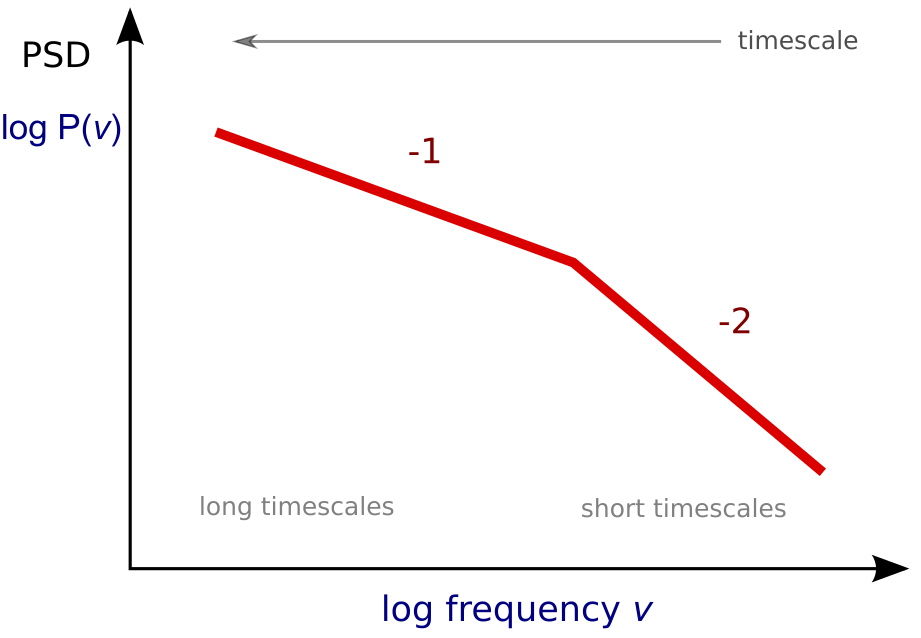}
 \caption{The PSD behavior of the VHE emission from PKS 2155-304 is suggestive of a break from flicker ($\beta=1$) to
 Brownian ($\beta=2$) power-law noise at a break frequency corresponding to a break time of $t_b\sim 1$ d.}\label{psd_model}
\end{figure}

PSD analysis has by now been performed for a variety of gamma-ray-emitting AGN, particularly in the high-energy
{Fermi}-LAT domain \cite{Chatterjee2012,Nakagawa2013,Sobolewska2014,Kushwaha2017,Goyal2018}. In the VHE
domain, a particularly interesting result relates to PKS 2155-304, where an extended PSD analysis revealed a flicker
noise behavior $\beta\sim1$ for its long-term low-level VHE activity over frequencies corresponding to timescales $\geq 1$ d
(and with a power index similar to the one seen at HE on timescales $\geq 10$ d), cf. Figure~\ref{PKS2155_PSD}, while the 2006 VHE
flaring state showed a red-noise behavior $\beta\sim 2$ (on frequencies corresponding to timescales $\leq 3$~h)~\cite{pks2155_2010,Abdalla2017}.
In the simplest interpretation, the PSD slope is considered to be stationary and to follow a power-law with a transition from
$\beta \sim1$ to $\beta\sim2$ at around $\nu_b\sim 1/t_b$, where $t_b \sim1$~d; see Figure~\ref{psd_model}. This would
then be comparable to the X-ray PSDs of Seyfert AGNs, which reveal a similar steepening by $\Delta \beta \simeq 1$ at their
break frequencies \cite{McHardy2006}. Interestingly, at X-ray energies, a break time of $t_b \sim1$~d has been suggested
earlier for PKS 2155-304 \cite{Kataoka2001}.
In principle, accretion-disk fluctuations (see above) could also account for a power-law noise behavior \cite{Lyubarski1997}.
In the case of the radio-quiet Seyfert AGNs, McHardy et al.~\cite{McHardy2006} have reported a simple quantitative relationship
between the break time $t_b$---then associated with the characteristic timescale at the inner edge of the accretion disk---the 
observed (bolometric) luminosity $L_B$, and the black hole mass $M_{\rm BH}$ of the source. Although PKS 2155-304
is not a radio-quiet object, one could speculate that a similar relation applies if the timing properties originate in the accretion
flow \cite{Abdalla2017}. For a standard disk (with the accretion rate as a proxy for $L_B$), this scaling relation becomes
$(t_b/1~\rm{d}) \simeq 0.7~(M_{\rm BH}/10^8 M_{\odot})^{1.12}/ \dot{m}_E^{0.98}$, where the accretion rate $\dot{m}_E$
has been expressed in units of the Eddington rate. Hence, if this relation also applies to the supposed VHE break time
of PKS 2155-304 ($t_b \sim 1$ d), accretion rates close to Eddington would be implied even for the quiescent VHE state.
It seems thus more likely that the break time in PKS 2155-304 is related to a change in accretion flow conditions such as a
transition from an inner advection-dominated to an outer standard disk configuration \cite{Abdalla2017}.

Obviously, Fourier analysis of non-thermal emission models could allow one to gain further insights into the origin of the gamma-ray
variability in blazars. This particularly includes recent explorations as to the possible modifications of an underlying (injected)
PSD shape by radiation \cite{Finke2014,Finke2015}. This can be done by starting from the appropriate time-dependent particle
transport equation (including the relevant acceleration and radiative loss terms) for $N_e(\gamma,t)$, the number of electrons
between $\gamma$ and $\gamma+d\gamma$ at time $t$. One can then look for solutions $\tilde{N}_e(\gamma,f)$ of the
Fourier-transformed equation assuming injection of power-law noise with $\tilde{Q}(\gamma, f) \propto f^{-\beta}$. This can
be used to, e.g., evaluate the impact on the expected synchrotron, external Compton (EC), or synchrotron self-Compton (SSC)
flux $\tilde{F}$ (with $\tilde{F}$ as the Fourier transform of the respective flux). Estimating the PSD from the periodogram, one,
e.g., finds that at sufficiently low frequencies $|\tilde{F}(SSC)(f)|^2 \propto f^{-(4\beta-2)}$ for SSC and $|\tilde{F}(EC)(f)|^2
\propto f^{-2\beta}$ in the case of EC \cite{Finke2014}. Since the gamma-ray emission in flat spectrum radio quasars (FSRQs)
and BL Lacs is thought to be dominated by different radiation processes (EC versus SSC), one might expect differences in
the PSD slopes for these sources if the injection ($\beta$) would be similar. In particular, if the above relation is applied to the BL
Lac object PKS 2155-304, believed to be dominated by SSC, this would suggest injection with $\beta\sim1$ (for the flare) and
$\beta \sim 0.75$ (for the long-term, low-level activity), respectively. The former could possibly be related to the flicker-noise
behavior induced by accretion disk fluctuations, while the latter would require some radial dependencies \cite{Lyubarski1997}.
At optical (R-band) frequencies, the emission of PKS 2155-304 has been reported to obey a PSD with slope $\sim$ (1.8--2.4)
on timescales larger than several days \cite{Chatterjee2012} (but see also \cite{Chevalier2019}). In the above approach,
the PSD for synchrotron emission roughly follows $f^{-2\beta}$ and steepens towards higher frequencies ($f \simgt 1/t_{\rm esc}$)
if escape is included \cite{Finke2014}. Given current uncertainties in PSD slope determination, it seems thus possible to relate
the different slopes at optical and HE/VHE to synchrotron and SSC radiation processes, respectively (cf. also \cite{Chevalier2019}).
More advanced modeling in this regard, however, seems desirable.

From a formal point of view (and also relevant to the below), it seems important to keep in mind that parameter estimation based
on light curve simulations need to be consistent in as much as they account for both the details of the PDF (e.g., log-normality)
and the PSD.\footnote{The widely employed Timmer and K\"onig algorithm \cite{Timmer1995} to simulate light curves assumes
a normal (Gaussian) stochastic process, and thus potentially introduces errors in parameter estimation for log-normal processes
\cite{Emma2013}. It remains to be studied how this affects the results given the quality of available gamma-ray data.}

\subsection{Quasi-Periodic Variability}
\begin{figure}[H]
 \centering
 \includegraphics[width=300pt]{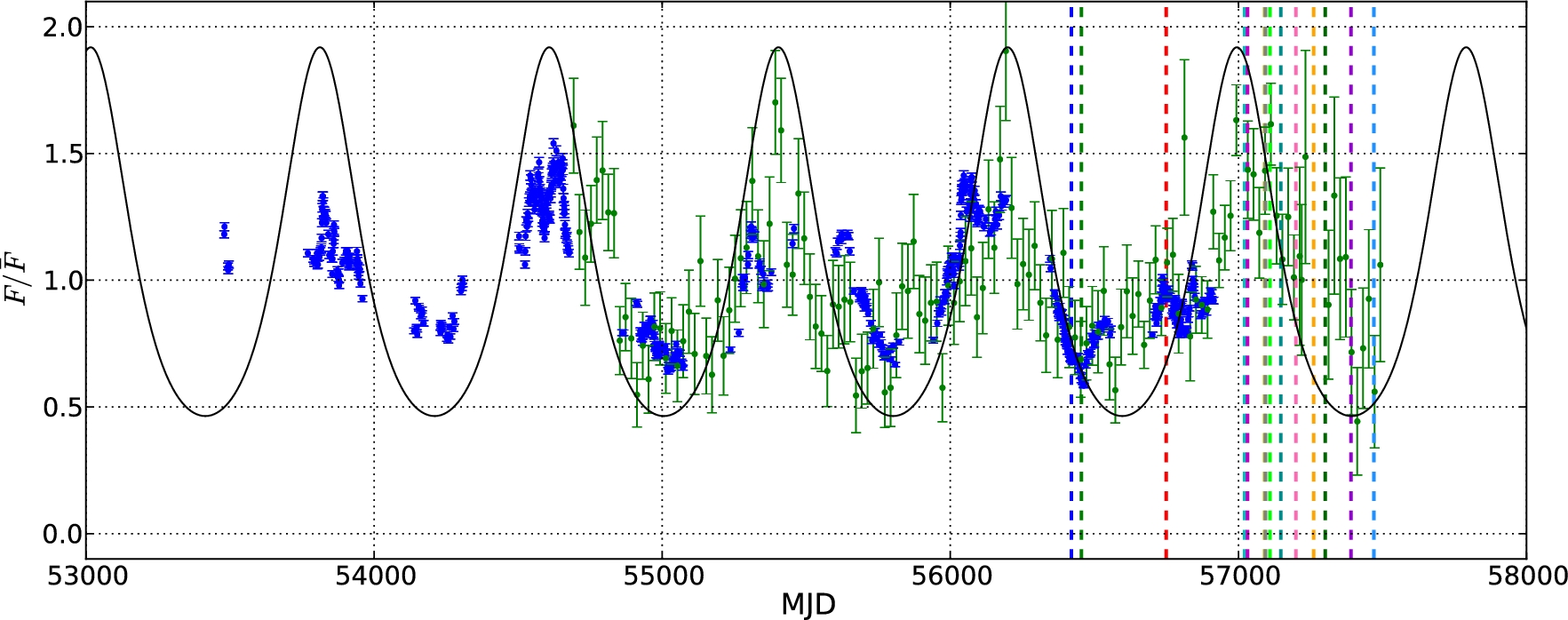}
 \caption{The $\sim9$ y-long light curve of PG 1553+113 in HE gamma-rays (20-d bins; green points) and in the
 optical (blue points) band. The black line shows the periodicity model, and the vertical line corresponds to some spectrum
 calculations. From~\cite{Sobacchi2017}.}\label{pg1553}
\end{figure}

With the availability of continuously-sampled gamma-ray light curves, periodicity analysis has gained new momentum as a
possible tool to probe deeper into the astrophysical nature of the sources. In particular, the (unbiased) {Fermi}-LAT light
curves of blazars have been used to search for quasi-periodic oscillations (QPOs) on yearly timescales. A prominent
example is provided by the BL Lac objects PG 1553+113 (at a redshift of $z\sim 0.5$), where a periodicity of $P=2.2$ y in
HE gamma-rays (and similarly, also in the optical) over $\sim 9$ y has been reported \cite{Ackermann2015,Sobacchi2017,
Tavani2018}; see Figure~\ref{pg1553} for an illustration.

Possible year-type HE periodicities (though at different significance levels) have now been reported for a variety of blazars
including PKS 2155-304 ($P \sim1.7$ y), Mkn 501 ($P\sim 0.9$ y), BL Lac ($P\sim 1.8$ y), PG 1553+113 ($P\sim 2.2$ y),
PKS 0426-380 ($P\sim 3.4$ y), PKS 0537-441 ($P\sim 0.8$ y during the high state), and PKS 0310-243 ($P\sim 2.1$ y), see
e.g.,~\cite{Ackermann2015,Sandrinelli2016,Sandrinelli2017,Prokhorov2017,Zhang2017,Zhang2017b,Bhatta2018}.

Multi-year quasi-periodic variability in AGN light curves has often been interpreted in the context of supermassive binary
black holes (SBBHs), such as the optical $\simeq12$-y periodicity in the BL Lac object OJ 287 ($z=0.3$) (e.g., \cite{Valtonen2016})
or the apparent optical $P\simeq5.2$-y periodicity in the quasar PG 1302-102 ($z=0.28$) \cite{Graham2015}.
In general, SBBHs are a natural stage of hierarchical galaxy formation in which elliptical galaxies (e.g., the host galaxies
of radio-loud AGN) are formed by galaxy mergers \cite{Begelman1980}. Depending on their evolutionary path, close SBBHs
(with separation less than a few parsecs) could potentially still reside in the center of (some) radio-loud AGN (e.g., \cite
{Komossa2006,Rieger2007,Merritt2013}). This is in fact supported by recent morphological evidence for SBBH-driven
geodetic precession ($P\sim$ $10^6$--$10^7$ y) in the radio maps of powerful AGN \cite{Krause2019}. Close, accreting SBBH
systems are likely to be accompanied by a circumbinary disk surrounding the binary, with circumbinary gas streams feeding
mini-disks around each black hole and affecting their evolution (e.g., \cite{Farris2014,Bowen2017}).

The HE findings noted above have led to the emergence of new SBBH interpretations. From a statistical point of view,
however, some caution needs to be exercised concerning the significance of the inferred periods, as in most cases
(perhaps apart from Mkn 501), only a few cycles are present over the available data. As shown by Vaughan et
al.~\cite{Vaughan2016}, clear phantom periodicities over $\sim$(2--3) cycles can be well found in pure noise data. Moreover,
there are indications that the QPO results may be more dependent on the analysis method employed than initially thought
\cite{Covino2019}. In terms of a global significance (assuming the absence of other physical reasoning) (cf., e.g., \cite
{Benlloch2001}), year-type QPOs are often no longer significant \cite{Faical2019}. Increasing the number of possible cycles
by continuous (well-sampled) monitoring will be important to better assess the current evidence and qualify its impact. In
addition, multi-wavelength comparison (as partly done for some of the sources) could improve the robustness of possible detections,
though (physically) periods may not necessarily have to be coinciding for very different wavebands. The HE {Fermi}-LAT
$\gamma$-ray observations could in principle be complemented by observations in the VHE domain, though gaps and uneven
sampling may represent a more serious issue. Combining different VHE instruments such as FACT and HAWC (e.g., 
see~\cite{Dorner2017} for the case of Mkn 501) could however improve the situation.

While speculative, one could still try to evaluate possible physical mechanisms for periodicity in AGN. While SBBHs are
expected, they may often not provide the best explanation for the noted QPOs: QPOs with physical periods $P_r = P/(1+z)
\sim 1$ y would imply very close binary systems and thus a significant amount of gravitational wave emission. This
would result in a very short gravitational lifetime of the system, $T \sim$ $(10^3$--$10^4)$ y, making it unlikely that we
should be able to detect many of such systems. Similarly, Pulsar Timing Array observations impose upper limits on the
(nano-hertz) gravitational stochastic background. If the binary hypothesis is tested for the HE blazar population, assuming
year-like QPOs and using the respective luminosity functions, only a very minor fraction (0.01--0.1\%) of BL Lacs and
FSRQs could harbor SBBHs \cite{Holgado2018}. In fact, pervious results are rather suggestive of orbital periods of the
order of $\sim 10$ y \cite{Rieger2007}. It seems thus more promising to relate year-type QPOs to other origins such
as the helical motion of a component inside a rotating jet, where differential Doppler boosting accounts for a periodic
lighthouse effects \cite{Camenzind1992}; or some quasi-periodic modulation in the accretion flow (e.g., induced by
modulations of the transition from an ADAF to a standard disk) feeding the jet (e.g., \cite{Gracia2003}). Constraints on
the jet radius suggest possible periods $P < 2$~y for the former scenario (cf., \cite{Rieger2004}), while the latter
would suggest transition radii $r_{tr} \sim 100 r_g$. At the current stage, it seems important to evaluate the adequacy
of such explanations in more detail.

We note that a very interesting, month-type ($34.5$ d) HE QPO over six cycles has been recently reported for the
BL Lac object PKS 2247-131 ($z=0.22$) \cite{Zhou2018}. The QPO behavior follows an outburst of the source in
October 2016 and is thought to be associated with a helical jet structure induced by the orbital motion in an SBBH.
The observed month-type HE QPO could in principle results from a shortening of the real physical period due to
relativistic travel-time effects (e.g., \cite{Rieger2004}), in which case it would be suggestive of a (physical) orbital
period of $P\simeq 7.1$ y \cite{Zhou2018}. Interestingly, a similar month-type ($\sim23$ d) behavior (over 4--5
cycles) at VHE energies has been reported for Mkn 501 during a large flare in 1997 \cite{Hayashida1998,Osone2006}
and has also been interpreted in an SBBH framework \cite{Rieger2000}.

\section{A Possible Example: PKS 2155-304}
\begin{figure}[H]
 \centering
 \includegraphics[width=380pt]{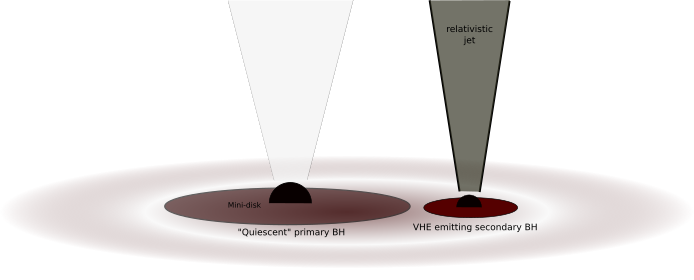}
 \caption{A possible binary black hole model for PKS 2155-304, with the binary (black holes plus mini-disks)
 being surrounded by a circumbinary disk. The tidal gas stream from this disk preferentially feeds the less massive,
 secondary black hole, whose jet emission dominates the observed VHE radiation spectrum. Both the short
 VHE variability and the high (apparent) luminosity can thus be accommodated, cf.~\cite{Rieger2010}.}\label{binary_model}
\end{figure}

Given the above-noted VHE timing characteristics of PKS 2155-304 (log-normality and power-law noise behavior)
and the properties of disk fluctuations, one could hypothesize that the VHE variability is driven by density fluctuations
in the accretion-disk. If these are efficiently transmitted to the jet, it could lead to a respective, power-law noise spectrum
in injection for Fermi-type particle acceleration. One~would only be able, however, to observe rapid VHE variations
(occurring on timescales as short as $\Delta t_{\rm obs} \simeq 200$ s) with the corresponding disk characteristics
if these signatures do not get obscured by other processes occurring on longer timescale within the source. As the
flux changes seen by an observer will appear to be convolved and thus dominated by the longest timescale, this
would require that the (observed) timescales for photons traveling across the radial width of the source and for the
relevant radiative loss processes still remain smaller than $\Delta t_{\rm obs}$. As shown in~\cite{Rieger2010},
this seems feasible in the case of PKS 2155-304. To account for the observed minimum variability, however, the
black hole mass would need to be limited to $m_{\rm BH} \leq 4 \times 10^7 M_{\odot}$ (assuming a thick inner disk);
see Equation~(\ref{BH_size}). In particular, the scenario would require a black hole mass that is seemingly smaller than the
total central black hole mass inferred from the $M_{BH} - L_{bulge}$ relation. A straightforward way to account for this
is to consider a binary black hole system where the jet dominating VHE emission is emitted from the secondary (less
massive) black hole \cite{Rieger2010}; see Figure~\ref{binary_model}. As noted above, viewing elliptical galaxies as
a merger product of spiral galaxies, an SBBH stage is expected to occur at some time. Once the secondary becomes
embedded in the outer disk around the primary (more massive) black hole, it will start clearing up an annular gap.
Mass supply from the circumbinary disk to the central binary, however, can still continue through tidal, time-dependent
(periodically-modulated) gas streams, which penetrate the gap and preferentially feed the secondary (disk) (e.g., \cite
{Artymowicz1996,Hayashida2007,Farris2014}). Numerical simulations in particular reveal that a reversal of mass
accretion rates can occur (i.e., $\dot{m}_{BH,2} > \dot{m}_{BH,1}$ despite $m_{BH,2} <m_{BH,1}$), acting towards
equal-mass binaries and suggesting that the secondary and its jet can become more luminous than the primary. In
such a scenario, a high VHE luminosity, as inferred for PKS 2155-304, could be accounted for, despite the smaller
(secondary) black hole mass.

When put in context, these considerations suggest that care needs to be exercised in the interpretation of VHE
variability constraints to avoid a spurious identification of the emitting (possibly secondary) black hole mass with
the total (possibly primary + secondary) central black hole mass of the system. In the case of PKS 2155-305, there
are additional pieces of circumstantial evidence, such as a (comparable) small black hole mass inferred from its
X-ray variability (PSD) properties (e.g., \cite{Czerny2001}), indications for optical long-term ($\sim7$ y) periodicity
\cite{Fan2000}, and jet bending on pc-scales (e.g., \cite{Piner2010}), which may in fact be more easily integrated
within a binary framework.

\section{Conclusions}
Time variability provides key information about the physics of astrophysical objects beyond that given solely in
their energy spectra. As indicated in this paper, this information is vital to properly understand the phenomena
and processes involved. Given the progress in gamma-ray astronomy, statistical tools can now be applied to
study the timing properties of AGN and to characterize their variability in terms of, e.g., flux distributions (PDF)
and (temporal) frequency dependence (PSD), along with characteristic timescales and possible QPOs.
This positive development is set to strengthen with upcoming instrumental developments and will make it
possible to deepen our understanding of the link between accretion dynamics, black hole physics, and jet
ejection.

\vspace{6pt}
\funding{Funding by a DFG Heisenberg Fellowship RI 1187/6-1 is gratefully acknowledged.}

\acknowledgments{Support and hospitality of the Max-Planck-Institut f\"ur Kernphysik (MPIK), Heidelberg, is 
gratefully acknowledged. I would like to thank Daniela Dorner and Thomas Bretz for organizing an inspiring 
conference, and Nachiketa Chakraborty for helpful comments on the manuscript. I thank S. Sahayanathan
and R. Khatoon as well as E. Sobacchi for their permission to use Fig. 7 and 10, respectively.}

%

\reftitle{References}

\end{document}